\documentclass[pra.aps,showpacs]{revtex4}
\usepackage{graphicx}
\usepackage{dcolumn}
\usepackage{bm}
\newcommand{\beq}{\begin{equation}}
\newcommand{\eeq}{\end{equation}}
\newcommand{\beqa}{\begin{eqnarray}}
\newcommand{\eeqa}{\end{eqnarray}}

\newcommand{\ra}{\rangle}

\begin{document}

\title{Convolutionless Non-Markovian
master equations and quantum trajectories: 
Brownian motion revisited}

\author{Walter T. Strunz$^1$\footnote{Electronic address: walter.strunz@physik.uni-freiburg.de}  
and Ting Yu$^{2,3}$\footnote{Electronic address: ting@pas.rochester.edu}}
\affiliation{ $^1$ Theoretische Quantendynamik, Physikalisches Institut,
Universit\"at Freiburg, Hermann-Herder-Str. 3, 79104 Freiburg,
Germany\\
 $^2$ Department of Physics 
and Astronomy,  University of Rochester, Rochester, NY14627, USA\\ 
 $^3$ Department of Physics, Queen Mary College, University of London, Mile End
Road,\\ London E1 4NS, United Kingdom}
%
\date{December 12, 2003}

\begin{abstract}
Stochastic Schr{\"o}dinger equations for quantum
trajectories offer an alternative and sometimes superior approach
to the study of open quantum system dynamics. Here we show that
recently established convolutionless non-Markovian stochastic 
Schr{\"o}dinger equations may serve as a powerful tool for the derivation of
convolutionless master equations for non-Markovian open quantum
systems. The most interesting example is quantum Brownian motion (QBM)
of a harmonic oscillator coupled to a heat bath of
oscillators, one of the most-employed 
exactly soluble models of open system dynamics.
We show explicitly how to establish the direct connection between 
the exact convolutionless master equation of QBM and the corresponding 
convolutionless exact stochastic Schr\"odinger equation.

\end{abstract}


\maketitle

\section{Introduction}\label{introduction}

Current attempts to push quantum coherent dynamics
further towards the macroscopic \cite{squid,zeili} or towards systems 
residing on more and more particles \cite{wineferd}
have to overcome decohering influences of the environment.
Typically, the system in question becomes entangled with
environmental degrees of freedom and thus coherence is lost.
Even if seemingly weak, couplings to an environment may have
dramatic consequences for coherences.
Thus, any theoretical modeling of
such quantum dynamics should take into account environmental effects.

Traditionally, open quantum systems are described in terms of
their reduced density operator and the dynamics thereof. 
Ideally, maps, propagators or even
evolution equations are derived that account for environmental effects
\cite{CAR,GZ,Weiss,CAR2,BPbook}.

The last decade has seen tremendous progress in the
development of stochastic Schr{\"o}dinger
equations to describe open quantum system dynamics 
\cite{CAR,GZ,DM,GP,GPZ,WM,PK}.
In this framework, the reduced density operator is 
obtained from an ensemble mean over the pure state solutions --
``quantum trajectories'' -- of a stochastic Schr\"odinger equation.
While known to exist in the case of
Markovian open system dynamics (when the ensemble dynamics
is governed by a Lindblad master equation \cite{LIN,GKS}),
it is only fairly recently
that quantum trajectories were extended to cover
more general non-Markovian open systems
\cite{S96,DS,DGS,DGS1,JCW,BP,YDGS,SDGY,CRE,Gambetta1}.
Interestingly, in our approach \cite{DS,DGS,DGS1} the non-Markovian
stochastic Schr\"odinger equation is derived directly from
an underlying microscopic system-plus-environment model without 
referring to the
corresponding evolution of the reduced density operator.
This turns out to be an efficient way to tackle non-Markovian 
quantum dynamics, shedding new light on the difficulties encountered
when dealing with memory effects.

Non-Markovian dynamics usually means
past contributions to the current time evolution: memory
effects typically enter through integrals over the past
\cite{NAK,ZWA}.
Nevertheless, under certain circumstances, non-Markovian
dynamics may be cast into a time-local, convolutionless form
where the dynamics of the open system state is determined by 
the state at the current time $t$ only \cite{ST64,vK,BKP}. 
Then, non-Markovian effects
are taken into account by certain time dependent coefficients
that may replace the memory integrals usually encountered.

Examples of this class are
models of open quantum system dynamics that may be treated without
any approximation. The famous damped harmonic
oscillator bilinearly coupled to a bath of
harmonic oscillators
\cite{FEY,ST64,Ull66,CAL,HR,GRA,UZ,HPZ,HPZ93} allows for
a convolutionless treatment without any approximation.
Such model systems allow us to test approximation schemes and, 
very often, they enable us to pin down important qualitative features 
of the dynamics of more realistic open quantum systems clearly.

The purpose of this paper is to show that our non-Markovian quantum
trajectory approach, besides its powerful use in numerical
simulations, provides an elegant way to derive
non-Markovian convolutionless master equations for open quantum systems. We
exemplify this idea by deriving the exact 
convolutionless non-Markovian master
equation for the famous quantum Brownian motion model of
a harmonic oscillator \cite{HR,HPZ}. 
This QBM master equation has been extensively studied in
different contexts 
\cite{FEY,ST64,Ull66,CAL,HR,GRA,UZ,HPZ,HPZ93,HY96,CF,PAZ,CRV,SH}. Our
approach is different from all the previous work.
We show in this paper how to establish the direct link
between our convolutionless stochastic Schr\"odinger equation and
the known convolutionless master equation.
The key ingredient for the derivation 
is a Heisenberg operator
approach to our non-Markovian stochastic Schr\"odinger equation \cite{CRE}.

Our paper is organized as follows: in Section II we review the
`` non-Markovian quantum state diffusion '' stochastic
Schr\"odinger equation, sketch its derivation
and display different approaches to a convolutionless
formulation. Next we show how to derive master equations from
the stochastic Schr\"odinger equation and give some simple
examples. The main part of the paper follows in Sections IV and V
where we discuss both the stochastic approach to the soluble
QBM-model and the derivation of the convolutionless master
equation based on the stochastic approach. Unavoidable
technical details are left for two appendices.

\section{Stochastic Schr\"odinger equation}
\label{mastervsschrodinger}

Here we review the Markov and, more importantly, our non-Markovian
stochastic approach to open quantum system dynamics \cite{DGS,DGS1}.
The dynamics of the reduced density operator $\rho_t$ is obtained from
a stochastic evolution equation for pure states $\psi_t$, such that
upon taking the ensemble mean ${\cal M}\{\ldots \}$
over the pure states,
the reduced density operator is recovered:
\begin{equation}\label{quensmean}
\rho_t = {\cal M}\{|\psi_t\rangle\langle\psi_t|\}.
\end{equation}

\subsection{Usual Markov limit: Lindblad theory}

In the standard Markov case, often encountered in quantum optical
applications, the evolution equation
for the reduced density operator of the open system
takes the Lindblad form \cite{LIN,GKS} 
\begin{equation}\label{lind}
\partial_t\rho_t = -\frac{i}{\hbar}[H,\rho_t]
+\frac{1}{2}\left([L,\rho_t L^\dagger]+[L \rho_t, L^\dagger]\right).
\end{equation}
We here restrict ourselves to the simplest 
case of a single non-unitary contribution 
involving a single operator $L$ (the {\it Lindbladian}).
In general, a sum or integral over such Lindbladians may appear.
Here and throughout the paper, we denote with $H$
the Hamiltonian that generates the otherwise isolated, unitary
evolution of the system.

A stochastic Schr\"odinger equation ``unraveling''
Lindblad evolution (\ref{lind}) is
provided by the linear {\it quantum state diffusion} \cite{GP,vK} equation
\begin{equation}\label{lqsd}
\partial_t\psi_t = \left(
-\frac{i}{\hbar}H
+ L z_t^* - \frac{1}{2}L^\dagger L\right)\psi_t
\end{equation}
for unnormalized stochastic states $\psi_t$.
The stochastic process $z_t^*$ is complex white noise
with zero mean (${\cal M}\{z_t^*\}=0$) and
correlations
\begin{equation}\label{zcorrdelta}
{\cal M}\left\{z_t z_s^*\right\} = \delta(t-s)\;\;\;\;\mbox{and}
\;\;\;\;
{\cal M}\left\{z_t z_s\right\} = 0.
\end{equation}
Equation (\ref{lqsd}) is here understood as a stochastic equation
in its
Stratonovich sense, though differences to the Ito case only
appear in their respective nonlinear versions, which are not
relevant for this paper, see \cite{GP}.
As one may easily confirm,
provided the quantum trajectories follow equation
(\ref{lqsd}), the ensemble mean (\ref{quensmean})
indeed evolves according to Lindblad's equation (\ref{lind}).

\subsection{Non-Markovian stochastic Schr\"odinger equation}

In order to extend the Markov theory we use a
standard model for open quantum systems in the
system-plus-reservoir framework:
a system with Hamiltonian $H$, coupled linearly to a large number 
of harmonic oscillators with distributed eigenfrequencies
$\omega_{\lambda}$ and creation and annihilation
operators $b^{\dagger}_{\lambda}, b_{\lambda}$. This total Hamiltonian
can be written as
\begin{equation}\label{totalH}
H_{\rm tot} =H + H_{\rm int} + H_{\rm bath}
= H + \hbar \sum_\lambda (g^*_\lambda
Lb_\lambda^\dagger  + g_\lambda L^\dagger  b_\lambda) +
\sum_\lambda \hbar\omega_\lambda b_\lambda^\dagger b_\lambda,
\end{equation}
where $L$ is the system operator providing the coupling to the environment, 
and $g_\lambda$ are coupling constants.

The interaction between system and 
environment may be written in the more appealing form
\begin{equation}\label{hint}
H_{\rm int} = \hbar(LB^\dagger +L^\dagger B)
\end{equation}
with a bath operator $B$ consisting of contributions of all
environmental oscillators,
$B= \sum_\lambda g_\lambda b_\lambda$.
It turns out to be convenient to change to an interaction representation
with respect to the free bath evolution, such that instead
of (\ref{totalH}) we
use
\begin{equation}\label{totalht}
H_{\rm tot}(t) \equiv e^{\frac{i}{\hbar}H_{\rm bath}t}(H
+H_{\rm int})e^{-\frac{i}{\hbar}H_{\rm bath}t} 
= H + \hbar\left(LB^\dagger(t)+L^\dagger B(t)\right)
\end{equation}
for the total Hamiltonian with 
\begin{equation}\label{bath}
B(t)=  e^{\frac{i}{\hbar}H_{\rm bath}t} B e^{-\frac{i}{\hbar}H_{\rm bath}t}
= \sum_\lambda g_\lambda b_\lambda e^{-i\omega_\lambda t}.
\end{equation}

The dynamics of the open system is thus influenced by the bath
operator $B(t)$, whose statistical properties are captured in its
correlation function
\begin{equation}\label{alpha}
\alpha(t-s) = \langle B(t) B^\dagger (s) \rangle_{\rm env} = 
\sum_\lambda |g_\lambda|^2 e^{-i\omega_\lambda(t-s)} = 
\hbar \int_0^\infty d\omega J(w) e^{-i\omega(t-s)},
\end{equation}
here for a zero temperature bath (for finite temperature see below). 
The last equality in (\ref{alpha}) defines the
spectral density $J(\omega)$ of bath oscillators.

If the dynamics of system and environment is such that
the bath correlation function in (\ref{alpha}) 
may be replaced by a delta function, the reduced dynamics
is Markovian and $\rho_t$ evolves according to the 
Lindblad master equation (\ref{lind}). For a general 
correlation function $\alpha(t-s)$, however,
memory effects of the environment may be important and
matters become exceedingly more difficult.
Such {\it non-Markovian} effects are known to be relevant in 
many situations, in particular at low temperatures and as soon
as narrow energy splittings occur, as in tunneling processes
\cite{Weiss}. They are also relevant for
structured environments, as for instance the electro-magnetic vacuum
in the presence of a photonic band gap material \cite{photobandgap}.
Moreover, non-Markovian effects play a role in output coupling
dynamics of atoms from a Bose Einstein condensate with the aim
to build an atom laser \cite{atomlaser}. 

Also, it may
happen that the dynamics appears to be Markovian, yet the
evolution equation is not of the standard Lindblad class (\ref{lind}).
This is the case, for instance, in the high-temperature limit
of the standard quantum Brownian motion model to be discussed 
in Sect. \ref{sqbm}.
In such cases, transient effects ({\it initial slips}, see also
\cite{fritz,gaspard1}) are
important and violate the semigroup property required for
Lindblad-class evolution.

The open quantum system described by the system-bath model
(\ref{totalH}), allows to derive the linear
non-Markovian generalization of (\ref{lqsd}),
henceforth referred to as the ``Non-Markovian quantum state
diffusion'' (NMQSD)- equation. The starting point is the formal
solution $|\Psi_t\rangle$ of the Schr\"odinger equation for the 
total system 
\begin{equation}
\label{schrequ} i\hbar\partial_t|\Psi_t\rangle
=H_{\rm tot}(t)|\Psi_t\rangle,
\end{equation}
in a special representation \cite{DGS}.
For simplicity, in this section we assume 
a zero temperature bath:
the initial pure state of system and environment is
\begin{equation}
|\Psi_0\rangle=|\psi_0\rangle\otimes |0_1\rangle\otimes
|0_2\rangle\cdots\otimes|0_\lambda\ra\otimes\cdots.
\end{equation}
The system state $|\psi_0\rangle$ is arbitrary and all environmental
oscillators
start in their respective ground state $|0_\lambda\rangle$.
By using a Bargmann coherent state basis
\cite{Barg} for the environmental degrees of
freedom:
$|z_{\lambda}\rangle=\exp\{z_{\lambda}a^{\dag}_{\lambda}\}|0_\lambda\rangle$,
and the resolution of the identity
\begin{equation}
I_{\lambda}=\int \frac{d^2 z_{\lambda}}{\pi}{\rm
e}^{-|z_{\lambda}|^2} |z_{\lambda}\rangle\langle z_{\lambda}|,
\end{equation}
the total state  $|\Psi_t\rangle$ in (\ref{schrequ}) can be expanded 
as
\begin{equation}\label{totalPsi}
|\Psi_t\rangle=\int \frac{d^2 z}{\pi}{\rm
e}^{-|z|^2}|\psi_t(z^*)\rangle\otimes|z\rangle,
\end{equation}
where $|z\rangle=|z_1\rangle\otimes |z_2\rangle\otimes \cdots
\otimes|z_{\lambda}\rangle \cdots$, $d^2z=d^2z_1d^2z_2\cdots
d^2z_\lambda\cdots$ and $|z|^2=\sum_\lambda |z_\lambda|^2$.
It follows from (\ref{totalPsi}) that the reduced density operator
of the open system may be obtained by an ensemble mean over
the system states $|\psi_t(z^*)\rangle$,
\begin{equation}\label{ensmeanproof}
\rho_t = \mbox{Tr}_{\rm bath}[|\Psi_t\rangle\langle\Psi_t|] =
\int \frac{d^2 z}{\pi}{\rm
e}^{-|z|^2} |\psi_t(z^*)\rangle\langle\psi_t(z^*)| \equiv 
{\cal M}\{|\psi_t\rangle\langle\psi_t|\}.
\end{equation}
The last equality defines the ensemble mean ${\cal M}\{\ldots\}$ as the
Gaussian integral 
$\int \frac{d^2 z}{\pi}{\rm
e}^{-|z|^2}\{\ldots \}$
over all coherent state labels $z_\lambda$.
Using the Schr\"odinger equation (\ref{schrequ}), the pure states 
$|\psi_t(z^*)\rangle$ of the system in (\ref{totalPsi})
were shown to satisfy the non-Markovian
QSD equation \cite{DS,DGS}
\begin{equation}\label{NMQSD}
\partial_t \psi_t = -\frac{i}{\hbar}H \psi_t + Lz^*_t \psi_t -L^\dagger
\int_0^t ds\;\alpha(t-s) \frac{\delta\psi_t}{\delta z^*_s}.
\end{equation}
where $\alpha(t-s)$ is the bath correlation function
(\ref{alpha}) and
$z_t^* = -i\sum_\lambda g_\lambda^*
z_\lambda^* e^{i\omega_\lambda t}$ is colored, complex Gaussian noise
with ${\cal{M}}\{z_t\}={\cal{M}}\{z_t z_s\}=0$ and 
${\cal M}\{z_t z_s^*\} = \alpha(t-s)$.

The linear non-Markovian QSD equation (\ref{NMQSD}) was here derived
for a zero-temperature heat bath. However, it can be shown
that equations (\ref{quensmean}) and (\ref{NMQSD})
remain valid
also at finite
temperatures provided we deal with Hermitian Lindbladians,
$L=L^\dagger$.
In general, the finite temperature non-Markovian QSD
equation takes a slightly more complicated form \cite{DGS,YU}.

\subsection{Convolutionless stochastic Schr\"odinger equation}

Memory effects of non-Markovian evolution clearly make themselves
felt through the integral over the past in equation (\ref{NMQSD}),
involving the bath correlation function and a functional derivative
of the current state $\psi_t(z^*)$ with respect to earlier
noise $z_s^*$. In many relevant cases, it is possible to
replace that functional derivative by some time dependent operator
$O$,
\begin{equation}\label{oop}
 \frac{\delta \psi_t(z^*)}{\delta z^*_s} = O(t,s,z^*)\psi_t(z^*),
\end{equation}
acting in the Hilbert space of the open system on the current
state $\psi_t(z^*)$ \cite{DGS}.
We indicate that $O$ may depend on the times $t$ and $s$,
and possibly on the (entire history of the) stochastic
process $z_t^*$. Relevant examples of this replacement
will be given shortly. One way to determine the
operator $O(t,s,z^*)$ in actual applications \cite{DGS} is to
insert Ansatz (\ref{oop}) in (\ref{NMQSD})
and use the ``consistency condition''
\begin{equation}\label{consistency}
\partial_t \frac{\delta\psi_t(z^*)}{\delta z^*_s} =
\frac{\delta}{\delta z^*_s} \partial_t\psi_t(z^*)
\end{equation}
together with the initial condition:
\begin{equation}
\label{initialcond}
O(t=s,s,z^*) = L
\end{equation}
for all $s$.

One finds the formal evolution equation
\begin{equation}\label{Oopevo}
\partial_t O(t,s,z^*) = \left[-{\frac{i}{\hbar}}H 
+ Lz_t^* - L^\dagger{\bar O}(t,z^*),
O(t,s,z^*)\right] - L^\dagger \frac{\delta {\bar O}(t,z^*)}{\delta z^*_s}
\end{equation}
with the time-integrated operator ${\bar O}(t,z^*)$:
\begin{equation}
\label{opbar}
{\bar O}(t,z^*)=\int^t_0 \alpha(t-s)O(t,s,z^*)ds.
\end{equation}
Equation (\ref{Oopevo}) has to be solved
with initial condition (\ref{initialcond}) for all $s$.
As has been shown in previous publications \cite{DGS}, 
one can determine the $O$-operator
in (\ref{oop}) for many interesting and 
relevant physical models. Moreover,
approximate $O$-operators can always
be found systematically and be easily implemented in
numerical simulations \cite{YDGS,SDGY,PS}.

Once the replacement (\ref{oop}) of the functional
derivative by an operator is known -- sometimes only approximately --
the NMQSD equation (\ref{NMQSD}) takes the more
useful {\it convolutionless} form \cite{DGS}
\begin{equation}\label{stoch3}
\partial_t \psi_t(z^*) =\left( -\frac{i}{\hbar}H
+ Lz_t^*-L^\dagger{\bar O}(t,z^*)
\right)
\psi_t(z^*),
\end{equation}
where we used the notation ${\bar O}(t,z^*)$ from (\ref{opbar}),
see also \cite{YDGS2}.
The determination of $\psi_t(z^*)$ is now reduced to solving
the simple time-local stochastic Schr\"odinger equation (\ref{stoch3}).
For a memory-less bath with $\alpha(t-s)=\gamma\delta(t-s)$, we see
from (\ref{initialcond}) and (\ref{opbar}) that
${\bar O}(t,z^*) = \frac{\gamma}{2}L$ holds. Then equation
(\ref{stoch3}) reduces indeed to the Markov linear quantum state
diffusion equation (\ref{lqsd}).

We recall that (\ref{stoch3}) does not
preserve the norm of the states $\psi_t(z^*)$. Thus, if
only a numerical solution of (\ref{stoch3}) is possible, it is 
most often more advisable to use its nonlinear, norm-preserving version
\cite{DGS}.

\subsection{Heisenberg approach}\label{heisenberg}

There is an alternative approach to the convolutionless 
NMQSD stochastic Schr\"odinger equation (\ref{stoch3}), see
\cite{CRE}. We find it most appropriate to work in terms of a
stochastic propagator $G_t(z^*)$ for the states $\psi_t(z^*)$.
By definition, we have
\begin{equation}
\label{recal} |\psi_t(z^*)\rangle=G_t(z^*)|\psi_0\rangle.
\end{equation}
We use the unitary propagator $U_t$ for the total state (\ref{totalPsi}),
satisfying
\begin{equation}\label{schrequop}
i\hbar\partial_t U_t=H_{\rm tot}(t)U_t\>  \> {\rm and } \> \> \>
U_0=I,
\end{equation}
with $H_{\rm tot}(t)$ from (\ref{totalht}).
Since $|\Psi_t\rangle = U_t|\Psi_0\rangle$ and 
$|\psi_t(z^*)\rangle = \langle z|\Psi_t\rangle$ according to
(\ref{totalPsi}),
the stochastic propagator $G_t(z^*)$ may be expressed as
\begin{equation}\label{stopro22}
G_t(z^*) = \langle z | U_t | 0 \rangle.
\end{equation}

We take the time derivative of the propagator in
(\ref{stopro22}) and obtain from the Schr\"odinger equation
(\ref{schrequop})
$i\hbar\partial_t G_t(z^*) = \langle z |H_{\rm tot}(t) U_t | 0
\rangle$. With Hamiltonian (\ref{totalH}), we arrive at
\begin{equation}\label{firstnewlnmqsd}
\partial_t G_t(z^*) = -\frac{i}{\hbar} H G_t(z^*) + L z_t^* G_t(z^*)
- i L^\dagger\sum_\lambda
g_\lambda e^{-i\omega_\lambda t} \langle z |b_\lambda U_t |0\rangle,
\end{equation}
with
$z_t^* = -i\sum_\lambda g_\lambda^* z_\lambda^* e^{i\omega_\lambda t}$
as before.

As in \cite{CRE}, we write
$\langle z | b_\lambda U_t |0\rangle=
\langle z | U_t b_\lambda(t) |0\rangle$ with the Heisenberg
operator $b_\lambda(t) = U_t^{-1} b_\lambda U_t$.
From the corresponding Heisenberg equation of motion,
\begin{equation}
i\hbar\partial_t b_\lambda(t) = U_t^{-1} [b_\lambda, H_{\rm tot}(t)] U_t =
\hbar g^*_\lambda e^{i\omega_\lambda t} L(t)
\end{equation}
with the Heisenberg operator $L(t) \equiv U_t^{-1}L U_t$, we find
upon integration
\begin{equation}\label{envaheisen}
b_\lambda(t) = b_\lambda - i g^*_\lambda\int_0^t ds\; L(s)
e^{i\omega_\lambda s}.
\end{equation}
We may conclude that
\begin{equation}\label{heisenberga}
\langle z | b_\lambda U_t |0\rangle= \langle z | U_t b_\lambda(t)
|0\rangle =-ig^*_\lambda\int_0^t ds\; e^{i\omega_\lambda s}
\langle z| U_t L(s)|0\rangle,
\end{equation}
where we have used $b_\lambda|0\rangle = 0$.

Replacing the last term in (\ref{firstnewlnmqsd}) by
expression (\ref{heisenberga}), the evolution equation for
the stochastic propagator becomes
\begin{equation}\label{secondnewlnmqsd}
\partial_t G_t(z^*) = -\frac{i}{\hbar} H G_t(z^*) 
+ L z_t^* G_t(z^*) -  L^\dagger
\int_0^t ds\;\alpha(t-s) \langle z|U_t L(s)|0\rangle,
\end{equation}
with the expected correlation function
$\alpha(t-s) = \sum_\lambda |g_\lambda|^2 e^{-i\omega_\lambda(t-s)}$,
at zero temperature.

Next it turns out appropriate to introduce the operator
\begin{equation}\label{capitalO}
{\cal O}(t,s) = U_t L(s) U_t^{-1} = U_t U_s^{-1} L U_s U_t^{-1}
\end{equation}
in the Hilbert space of system and environment. It allows us to
express the term under the memory integral in (\ref{secondnewlnmqsd})
in the form
\begin{equation}\label{newo0}
\langle z|U_t L(s)|0\rangle
= \langle z|{\cal O}(t,s)U_t |0\rangle.
\end{equation}
Comparing equation (\ref{secondnewlnmqsd}) with the NMQSD equation
(\ref{NMQSD}) with replacement (\ref{oop})
and recalling equation (\ref{stopro22}), we
see that we aim to find an operator $O(t,s,z^*)$ satisfying
\begin{equation}\label{newo}
\langle z|{\cal O}(t,s)U_t |0\rangle
= O(t,s,z^*)\langle z| U_t| 0\rangle
 = O(t,s,z^*)G_t(z^*).
\end{equation}
Relation (\ref{newo}) is crucial for finding an evolution
equation for $O(t,s,z^*)$ with respect to $s$, as to be shown 
in Section \ref{sqbm}. With expression (\ref{newo}) it is clear that
equation (\ref{secondnewlnmqsd}) is the linear
convolutionless non-Markovian stochastic Schr\"{o}dinger equation
(\ref{stoch3}).

Thus, the operator $O(t,s,z^*)$ in (\ref{newo}) (or (\ref{oop}))
may be found by investigating the equation of motion for the
{Heisenberg} coupling operator ${\cal O}(t,s) = U_t U_s^{-1} L U_s
U_t^{-1}$ in the Hilbert space of system and environment.
Depending on the application, this approach may offer a more
transparent way to establish an expression for the
$O(t,s,z^*)$-operator of a convolutionless formulation
than the evolution equation (\ref{Oopevo}). In
particular, it allows us to derive evolution equations for
$O(t,s,z^*)$ with respect to $s$, rather than $t$.

\section{Master equation}

Our stochastic approach to open quantum systems will be employed
to derive the master equation for the ensemble evolution \cite{YDGS,YDGS2}.
A particularly simple and straightforward route to the master
equation is possible as soon as the
operator $O(t,s,z^*)$ in (\ref{oop}) (or (\ref{newo})) and thus the
integrated operator ${\bar O}(t,z^*)$ from (\ref{opbar})
turns out to be independent of the noise $z^*$. This happens to be true for
many interesting cases, as shown below.
We stress, however, that for the main part of this paper, 
the Brownian motion model of Section \ref{sqbm},
${\bar O}(t,z^*)$ {\it does} depend on the noise process $z_t^*$ and
matters are more difficult.
 
\subsection{Density operator evolution}

According to our construction, the reduced density operator $\rho_t$
is given by the ensemble mean over the solutions of the
stochastic Schr\"odinger equation (\ref{stoch3}). We write
$\rho_t = {\cal M}\{P_t\}$ with
$P_t = |\psi_t(z^*)\rangle\langle\psi_t(z^*)|$.
Upon taking the time derivative in (\ref{quensmean})
and employing (\ref{stoch3}), we get
\begin{equation}\label{rhoevo}
\dot\rho_t = -\frac{i}{\hbar}[H,\rho_t] +
L{\cal M}\{z_t^*P_t\}-
L^\dagger{\cal M}\{{\bar O}(t,z^*)P_t\}
+{\cal M}\{P_tz_t\}L^\dagger-
{\cal M}\{P_t{\bar O}^\dagger(t,z^*)\}L.
\end{equation}
Apparently, this is still far from being a closed evolution equation for 
$\rho_t$.
Using a version of Novikov's theorem \cite{novikov},
\begin{equation}\label{novi}
{\cal M}\{P_t z_t\} = {\cal M}\left\{\int_0^tds\;\alpha(t-s)
\frac{\delta}{\delta z_s^*} P_t\right\},
\end{equation}
which amounts to a partial integration under the Gaussian
probability distribution in (\ref{ensmeanproof}),
we may use the replacement of the functional derivative
by the operator $O$ in (\ref{oop}) and find 
${\cal M}\{P_t z_t\} = {\cal M}\left\{{\bar O}(t,z^*)P_t\right\}$.
Then equation (\ref{rhoevo}) takes the form
\begin{equation}\label{rhoevo2}
\dot\rho_t = -\frac{i}{\hbar}[H,\rho_t] +
[L,{\cal M}\{P_t{\bar O}^\dagger(t,z^*)\}]+
[{\cal M}\{{\bar O}(t,z^*)P_t\},L^\dagger].
\end{equation}
A convolutionless evolution equation for
the non-Markovian dynamics of
the reduced density operator may be derived from the knowledge
of the operator $O(t,s,z^*)$ 
in (\ref{oop}). For instance,
as soon as the exact $O(t,s,z^*)$ is independent of
the noise $z_t^*$  or as soon as a noise-dependent $O(t,s,z^*)$ can 
be approximated well by a noise-independent operator $O(t,s)$, we find
$O(t,s,z^*)=O(t,s)$ and therefore
${\bar O}(t,z^*) = {\bar O}(t)$. In these cases,
equation (\ref{rhoevo2}) is a 
convolutionless closed master equation for $\rho_t$: due to
${\cal M}\{{\bar O}(t)P_t\} =
{\bar O}(t) {\cal M}\{P_t\}= {\bar O}(t) \rho_t$,
we find
\begin{equation}\label{evorho3}
\dot\rho_t = -\frac{i}{\hbar}[H,\rho_t] +
[L,\rho_t{\bar O}^\dagger(t)]+
[{\bar O}(t)\rho_t,L^\dagger].
\end{equation}
Let us emphasize that equation (\ref{evorho3}) is not of Lindblad form in 
general, not even in the long-time limit due to its non-Markovian
nature. Nevertheless, as long as we deal
with {\it exact} master equations -- as for most of this paper -- positivity
issues do not appear. By construction, despite their apparent non-Lindblad 
form, these exact non-Markovian convolutionless master equations represent
(completely) positive maps. However, as soon as the master equation
follows from an {\it approximate} replacement
of the operator $O(t,s,z^*)$, positivity is a difficult and 
delicate subject well beyond the scope of this paper. 
For such investigations on positivity
(importance of initial slips etc.) the interested 
reader is referred to the literature \cite{fritz,YDGS,gaspard1,YDGS2,SDGY}.

Clearly, the convolutionless master equation (\ref{evorho3}) covers a much 
wider class of open system dynamics than the Lindblad equation (\ref{lind}). 
The non-Markovian properties are encoded in 
a finite width correlation function $\alpha(t-s)$, and so in time-dependent 
coefficients appearing in the master equation.  By
construction, we already know the corresponding stochastic
Schr\"odinger equation (\ref{stoch3}).
Remarkably, here we {\it derive} the master
equation from the stochastic Schr\"odinger equation.

We stress that the master equ. (\ref{evorho3}) holds for a
noise independent operator $O(t,s,z^*)$ only.
Whenever $O(t,s,z^*)$ {\it does} depend on the noise $z^*$,
equ. (\ref{rhoevo2}) is still valid yet the step to equ. (\ref{evorho3})
fails and
matters are considerably more difficult. In this case, no general closed
evolution equation for $\rho_t$ is known.
However, starting from (\ref{rhoevo2}), one
may still be able to derive an evolution equation
for specific cases, as shown in this paper:
an important example of this class is
provided in Section \ref{sqbm} on Brownian motion,
where $O(t,s,z^*)$ depends linearly on the 
stochastic process $z_t^*$.

\subsection{Convolutionless master equations: simple examples}

As noted before, the operator $O(t,s,z^*)$ 
appearing in (\ref{oop}) is generally dependent on the noise $z^*$, so 
there is no general recipe how to derive the corresponding
master equation (see our approach in Sec. IV). Before proceeding to our 
main task, in what follows, we review a few examples where 
either the operator $O(t,s,z^*)$ according to (\ref{oop}) may be 
approximated by a noise-independent operator ${\bar O}(t)$ or the 
{\it exact} $O(t,s,z^*)$ does not contain the noise $z_t^*$. In both 
these cases, the corresponding convolutionless master equation 
(\ref{evorho3}) is valid.

\subsubsection{Weak coupling}

In many interesting cases, the assumption that the interaction between system
and environment is weak is a good approximation. 
The action of the functional derivative 
in (\ref{oop}) may be systematically expanded in
powers of the interaction $H_{\rm int}$
\cite{YDGS,gaspard2,PS}. 
\beq
O(t,s,z^*)=\sum_{n=0}^{\infty} g^nO_n(t,s,z^*)
\eeq
where $g$ represents the coupling strength. To lowest order,
one finds 
\beq
{\delta\psi_t(z^*)}/{\delta z_s^*} =
\left[e^{-iH(t-s)/\hbar}Le^{iH(t-s)/\hbar} + \ldots\right]
\psi_t(z^*)
\eeq
and thus in this lowest order the operator
\begin{equation}\label{perturboop}
O(t,s,z^*) \approx O_0(t,s,z^*) =
O_0(t,s) = e^{-iH(t-s)/\hbar}Le^{iH(t-s)/\hbar},
\end{equation}
independent of the noise. The relevant time-integrated expression
reads
\begin{equation}\label{perturbbaroop}
{\bar O}(t) \approx
{\bar O}_0(t) = \int_0^t ds\;\alpha(s)
e^{-iHs/\hbar}Le^{iHs/\hbar},
\end{equation}
and enters the master equation (\ref{evorho3})
for the ensemble evolution. Indeed, the resulting
equation is nothing but the weak coupling so-called ``Redfield''
master equation \cite{gaspard1,gaspard2},
including an ``initial slip'' captured by the initial time 
dependence of ${\bar O}(t)$. 
A weak coupling stochastic
Schr\"odinger equation equivalent to (\ref{stoch3}) (with
replacement (\ref{perturbbaroop})) was derived in \cite{gaspard2}
in a formulation that kept the memory integral
over the bath correlation function. The convolutionless
formulation (\ref{stoch3}) is easier to handle, as there is
no need to store the state vector $\psi_s(z^*)$ at earlier times
$s<t$.
 
\subsubsection{Near Markov}
If the bath correlation function $\alpha(t-s)$ falls off
rapidly under the memory integral in (\ref{NMQSD}),
an expansion of the functional derivative in terms of
the time delay $(t-s)$ is sensible \cite{YDGS,SDGY,YDGS2},
With $A_n(t) = \int_0^t ds\;s^n\alpha(s)$ and $n=0,1,2,\ldots$, 
we find for the relevant integrated operator
(\ref{opbar}) to first order,
\begin{equation}\label{markovbaroop}
{\bar O}(t) \approx A_0(t)L + A_1(t)\left(-\frac{i}{\hbar}
[H,L] + A_0(t)[L,L^\dagger]L\right),
\end{equation}
neglecting contributions $A_n(t)$ with $n\ge 2$. 
Once again, to the order accepted, we see the operator $O(t,s,z^*)$
turns out to be independent of the noise $z_t^*$ and the
master equation (\ref{evorho3}) immediately applies (see also \cite{YDGS2}).
 
Of particular interest is the standard Markov limit:
the correlation function may be replaced by a delta function,
$\alpha(t-s)=\gamma\delta(t-s)$, with some constant $\gamma$.
The only relevant term in (\ref{markovbaroop}) is the $A_0$-term which
may be replaced by the constant 
${\bar O} = \frac{\gamma}{2}L$. As noted earlier, 
the stochastic Schr\"odinger equation (\ref{stoch3})
(and thus, (\ref{NMQSD})) reduces to the linear version
of the Markov quantum state diffusion equation (\ref{lqsd}).
Correspondingly, the master equation (\ref{evorho3}) is nothing
but Lindblad's equation (\ref{lind}).
The next order $A_1$-term in expansion (\ref{markovbaroop})
turns out to be relevant for the high temperature limit
of the quantum Brownian motion model \cite{SDGY} and quite generally
leads to a theory of ``post-Markov'' evolution \cite{YDGS2}.

We remark that very often, Markov and weak coupling
approximation are only meaningful in combination,
referred to as {\it Born-Markov} approximation.
 
\subsubsection{Soluble models}
For some choices of system Hamiltonian $H$ and
coupling operator $L$ in the interaction Hamiltonian (\ref{hint}),
the functional derivative may be replaced by an operator $O(t,s,z^*)$
without any approximation \cite{DGS}. Two examples are mentioned,
where the $O$-operator turns out independent of the noise $z^*$. 
The third, more complicated case 
of Brownian motion of a harmonic oscillator will be the main 
subject of the following Sections.
 
The first simple example is an harmonic oscillator with
$H=\Omega a^\dagger a$, coupled to the environment
through a rotating-wave-type coupling, $L=a$ in the total
Hamiltonian (\ref{totalH}). It turns out \cite{DGS} that the functional
derivative in (\ref{oop}) 
may be replaced by an operator $O(t,s)$ being
proportional to the annihilation operator without any
approximation,
\begin{equation}\label{harmonoop}
O(t,s) = \frac{c(s)}{c(t)}a.
\end{equation}
Here, $c(t)$ is a complex function satisfying the
equation of motion
\begin{equation}\label{ceqmot}
\dot c(t) + i\Omega c(t) + \int_0^tds\;\alpha(t-s) c(s) = 0,
\end{equation}
involving the bath correlation function. We recognize
the damped motion of the amplitude of the oscillator.
The integrated operator ${\bar O}(t)$ thus becomes
\begin{equation}\label{osciintoop}
{\bar O}(t) = C(t)a
\end{equation}
without any approximation,
where $C(t) = \int_0^t ds\; c(s)\alpha(t-s)/c(t)$.
The resulting exact convolutionless
master equation \cite{lawande} thus follows from the general 
expression (\ref{evorho3}):
\begin{eqnarray}\label{evorho4}
\dot\rho_t & = & -i[\Omega a^\dagger a,\rho_t] + C^*(t)
[a,\rho_t a^\dagger]+ C(t)
[a\rho_t,a^\dagger] \\ \nonumber
& = & 
 -i[\left(\Omega+\mbox{Im}\{C(t)\}\right) a^\dagger a,\rho_t] 
+ \mbox{Re}\{C(t)\}
\left(
[a,\rho_t a^\dagger]+
[a\rho_t,a^\dagger]\right) \\ \nonumber
\end{eqnarray}

A similar result may be derived for a damped
two-level system with Hamiltonian $H = \Omega\sigma_z/2$
and the rotating-wave type coupling $L=\sigma_-$ \cite{DGS}. Similar to
the case of the damped (rotating-wave) harmonic oscillator above, 
one finds
\begin{equation}\label{twoleveloop}
O(t,s) = \frac{c(s)}{c(t)}\sigma_-
\end{equation}
with the very same function $c(t)$ from (\ref{ceqmot}).
Thus, an exact convolutionless master equation (\ref{evorho3})
of a form similar to (\ref{evorho4}) follows, with
$a$ replaced by $\sigma_-$ in the non-unitary part of the evolution equation.
 
To summerize, in all the cases presented in this
subsection it is possible to replace the functional derivative 
in (\ref{oop}) by an operator
$O(t,s)$ that is independent of the noise $z_t^*$.
In such cases, the ensemble mean follows the convolutionless master 
equation (\ref{evorho3}).
In the weak coupling
case with ${\bar O}(t)$ from (\ref{perturboop}),
for instance, we recover the so-called Redfield
master equation \cite{gaspard1,gaspard2}
including an initial slip. In the Markov case, keeping only
the lowest order in expansion (\ref{markovbaroop}),
we recover Lindblad's equation (\ref{lind}) for
the ensemble evolution. For the two exactly soluble
models with $O(t,s)$ from (\ref{harmonoop})
or (\ref{twoleveloop}), equation (\ref{evorho3}) is
the exact convolutionless master equation. In these latter cases,
asymptotically for large times $t$, when the
time dependent function $C(t)$ approaches a constant, 
the exact master equation turns into one of the
Lindblad class (\ref{lind}).
In the following Sections we discuss yet another exactly
soluble case, where the convolutionless 
master equation is more involved due
to the dependence of the operator $O(t,s,z^*)$ in (\ref{oop})
on the noise $z_t^*$. 

\section{Quantum Brownian motion}\label{sqbm}

\subsection{Brownian motion model}

Brownian motion of a harmonically bound particle
may be obtained from the Hamiltonian
\begin{equation}\label{qbmH}
H_{\rm tot} = \left(\frac{p^2}{2M} + \frac{1}{2}M\Omega^2 q^2\right)
+ q \sum_\lambda G_\lambda q_\lambda
+ \sum_\lambda\left\{ \frac{p_\lambda^2}{2m_\lambda} 
+\frac{1}{2}m_\lambda\omega_\lambda^2 q_\lambda^2 \right\}
\end{equation}
of an oscillator with mass $M$ and frequency $\Omega$,
coupled to an environment of harmonic oscillators through its 
position $q$
\cite{FEY,ST64,Ull66,CAL,HR,GRA,UZ,HPZ,HPZ93,HY96,CF,PAZ,CRV,SH}.
In this form, the usual ``counterterm'' arising from
the coupling should be
understood to be included in the harmonic potential, see
\cite{Weiss} for more details.

The Brownian motion Hamiltonian
(\ref{qbmH}) is of the form (\ref{totalH}) and thus
may be treated with
our stochastic approach to open quantum systems.
With $q_\lambda = \sqrt{\frac{\hbar}{2m_\lambda\omega_\lambda}}
(b_\lambda + b_\lambda^\dagger)$ we identify the 
quantities entering the basic Hamiltonian
(\ref{totalH}) to be $g_\lambda = G_\lambda
\sqrt{\frac{\hbar}{2m_\lambda\omega_\lambda}}$ and
the coupling operator is $L=L^\dagger=q/\hbar$.
The system Hamiltonian is the harmonic
\begin{equation}\label{newH}
H =  \frac{p^2}{2M} + \frac{1}{2}M\Omega^2 q^2.
\end{equation}

For the quantum bath correlation function (\ref{alpha})
of this model at temperature $T$ one finds
the force-force correlation
\begin{eqnarray}\label{alphatsTqbm}
\alpha(t-s) & = &  \left\langle(B(t)+B^\dagger(t))(B(s)+B^\dagger(s))
\right\rangle_{\rm env} \\ \nonumber
& = & \sum_\lambda \frac{\hbar G_\lambda^2}{2m_\lambda\omega_\lambda}
\left[\coth\left(\frac{\hbar\omega_\lambda}{2k_BT}\right)
\cos \omega_\lambda (t-s)
-i\sin\omega_\lambda (t-s) \right].
\end{eqnarray}

Introducing a spectral density of bath oscillators 
\begin{equation}\label{spectral}
J(\omega) = \sum_\lambda\frac{G_\lambda^2}{2m_\lambda\omega_\lambda}
\delta(\omega-\omega_\lambda)
\end{equation}
one usually writes
\begin{equation}\label{alphaqbm2}
\alpha(t-s) = \hbar\int_0^\infty d\omega\;J(\omega)
\left[\coth\left(\frac{\hbar\omega}{2k_B T}\right)\cos \omega (t-s)
-i\sin\omega (t-s) \right].
\end{equation}
Often the so-called Ohmic case is chosen with 
$J(\omega) = M\gamma\omega f_c(\omega/\Lambda)$ with some cutoff function
$f_c(x)$, e.g. $f_c(x) \simeq e^{-x}$ and $\Lambda$ a cutoff frequency.
We stress, however, that the following results are valid for any spectral
density $J(\omega)$.

\subsection{Master equation}
It is well known that
the model (\ref{qbmH}) allows the derivation of an exact convolutionless
master equation for the reduced density operator
\cite{HR,UZ,HPZ}. It may be written in the form
\begin{equation}\label{masterosci}
\dot\rho_t = \frac{1}{i\hbar}[H,\rho_t] 
+\frac{a(t)}{2i\hbar}[q^2,\rho_t] + \frac{b(t)}{2i\hbar}[q,\{p,\rho_t\}]
+\frac{c(t)}{\hbar^2}[q,[p,\rho_t]] - \frac{d(t)}{\hbar^2}[q,[q,\rho_t]]
\end{equation}
with real time dependent coefficients $a(t),b(t),c(t)$ and $d(t)$.
The physical meaning of these terms 
(drift terms $a(t)$ (frequency renormalization), $b(t)$ (damping term)
and diffusion terms $c(t), d(t)$) becomes apparent from
the Wigner representation of equation (\ref{masterosci}),
for which the
reader is referred to the literature \cite{HR,HPZ,HY96,CF,SH}.

In \cite{HR} the derivation of (\ref{masterosci}) 
was based on the generator of the time evolution, while
in \cite{HPZ,HPZ93}, the authors used path integrals.
In Section \ref{sqbm2} we show how the master equation 
(\ref{masterosci}) follows directly from the
stochastic Schr\"odinger equation which we discuss next.

\subsection{Stochastic Schr\"odinger equation}

Our stochastic Schr\"odinger equation approach allows a 
rigorous treatment of model (\ref{qbmH}). The replacement of
the functional derivative with an operator $O(t,s,z^*)$ 
in (\ref{oop}) can be established without any approximation,
and thus a convolutionless exact stochastic Schr\"odinger
equation (\ref{stoch3}) may be found. Here,
however, we have to allow for an explicit noise dependence.
It turns out \cite{DGS} that the ansatz
\begin{equation}\label{Oharmonic}
O(t,s,z^*) = \frac{1}{\hbar}\left(
f(t,s) q + \frac{1}{M\Omega}g(t,s) p - 
\frac{i}{M\Omega} \int_0^t ds' j(t,s,s') z_{s'}^*\right),
\end{equation}
with complex functions $f(t,s)$, $g(t,s)$, and $j(t,s,s')$ to
be determined is a solution of the general evolution
equation (\ref{Oopevo}).
We learn from this example that in general, the functional derivative
in (\ref{oop})
may indeed introduce a dependence on the whole history of the noise
$z_t^*$. 

The functions $f(t,s), g(t,s)$ and $j(t,s,s')$ in (\ref{Oharmonic})
have to satisfy the evolution equations
\begin{eqnarray}\label{fgjs}
\partial_t f(t,s) & = & \Omega g(t,s) - 2ig(t,s)F(t)+if(t,s)G(t)
                   +iJ(t,s), \\ \nonumber
\partial_t g(t,s) & = & -\Omega f(t,s) - ig(t,s)G(t),\;\;\mbox{and}\\ \nonumber
\partial_t j(t,s,s') & = & -ig(t,s)J(t,s'),
\end{eqnarray}
where we introduced the integrated functions
\begin{eqnarray}\label{FGJs}
F(t) & = & \frac{1}{M\Omega\hbar}\int_0^t ds\, \alpha(t-s) f(t,s),\\ \nonumber
G(t) & = & \frac{1}{M\Omega\hbar}\int_0^t ds\, \alpha(t-s) g(t,s),
         \;\;\mbox{and}\\ \nonumber
J(t,s') & = & \frac{1}{M\Omega\hbar}\int_0^t ds\, \alpha(t-s) j(t,s,s').
\end{eqnarray}
While $f,g$, and $j$ are dimensionless complex functions, the integrated
expressions $F,G$, and $J$ are defined such as to have dimension inverse
time: typically, these latter functions turn out to be proportional
to a damping rate $\gamma$.
The evolution equations (\ref{fgjs}) have to be solved with boundary
conditions 
\begin{eqnarray}\label{boundaries}
f(t=s,s) & = & 1,\;\;g(t=s,s)=0,\;\; j(t=s,s,s') = 0,\;\;\mbox{and}
\\ \nonumber
j(t=s,s',s) & = & -g(s,s'),
\end{eqnarray}
for all $s$ and $s'$.
The first three conditions arise from the
initial condition (\ref{initialcond}) for the operator $O(t=s,s,z^*)=L=q/\hbar$,
the last condition arising from the solution of (\ref{Oopevo}).
The relevant integrated operator (\ref{opbar})
turns out to be
\begin{equation}
{\bar O}(t,z^*) = M\Omega F(t)q + G(t) p -i\int_0^tds J(t,s) z_s^*
\end{equation}
with time dependent coefficients from (\ref{FGJs}).

The resulting linear, convolutionless non-Markovian
stochastic Schr\"odinger equation (\ref{stoch3}) for
the Brownian motion of a harmonic oscillator (\ref{newH})
with a coupling
to the environment through position $L=q/\hbar$ thus reads
\begin{eqnarray}\label{qbmosci}
\hbar \partial_t \psi_t & = & -iH\psi_t
 + q\left(z_t^* -{\bar O}(t,z^*)\right)\psi_t \\ \nonumber
& = & -iH\psi_t
 + q\left(z_t^* -M\Omega F(t) q -G(t) p +
i\int_0^t ds\;J(t,s) z_s^* \right)\psi_t.
\end{eqnarray}
The process $z_t^*$ is
a complex Gaussian process with the finite
temperature correlation function $\alpha(t-s)$
from (\ref{alphaqbm2}).
Just like the exact master equation (\ref{masterosci}),
the exact stochastic NMQSD
equation (\ref{qbmosci}) is valid for
arbitrary bath correlation function $\alpha(t-s)$, and thus for any 
temperature, environmental spectral density, or coupling strength
between the harmonic oscillator and its environment.

This example demonstrates clearly that the evolution
of quantum trajectories $\psi_t(z^*)$ may
indeed depend on the whole history of the noise.
Simulations of the Brownian motion model discussed here 
may be found in \cite{Str01}.  

From our construction, it is clear that the ensemble of
quantum trajectories of the exact equation (\ref{qbmosci}) evolves
according to the exact master equation (\ref{masterosci}).
In the next Section \ref{sqbm2} we set out to establish
this connection directly.

As ${\bar O}(t,z^*)$ depends
on the noise, the simple result (\ref{evorho3})
for the evolution of the reduced density operator
$\rho_t$ is not applicable, and further
effort is required. As a first step, we show in the next Section
how a Heisenberg operator approach helps to gain more insight
into result (\ref{Oharmonic}) for the operator
$O(t,s,z^*)$ replacing the functional
derivative.

\subsection{Heisenberg method}\label{heisenbergQBM}

Major clarification is achieved once we investigate the dependence
of the functions $f(t,s), g(t,s)$, and $j(t,s,s')$
in (\ref{Oharmonic}) on $s$ rather than $t$. The key step
arises from the Heisenberg operator approach as explained in
Section \ref{heisenberg}, see also \cite{CRE}.
Here, the challenge is to find a suitable expression for 
\begin{equation}\label{Aop2}
Q(s) \equiv \langle z|U_t q(s)|0\rangle 
        =\hbar O(t,s,z^*) \langle z|U_t|0\rangle,
\end{equation}
similar to expressions (\ref{newo0}) and (\ref{newo}).
Clearly, $Q(s)$ depends on the time $t$, but we here regard it as
a function of $s$ and thus may investigate the $s$-dependence of
the desired operator $O(t,s,z^*)$.
Not surprisingly, the Heisenberg equation of motion
for $q(s)$ leads to 
$\partial_s Q(s) = \langle z| U_t \partial_s q(s)|0\rangle
 = \frac{1}{M}\langle z| U_t p(s)|0\rangle$
and therefore we investigate the second order derivative.
Note that required final values are
\begin{eqnarray}\label{finalAs}
Q(s=t) & = & q \langle z| U_t|0\rangle \\ \nonumber
\left.\partial_sQ(s)\right|_{s=t} & = & \frac{1}{M}
p \langle z|U_t|0\rangle.
\end{eqnarray}
For the second order derivative
we find from (\ref{qbmH}) with 
$g_\lambda = \sqrt{\frac{\hbar}{2m_\lambda\omega_\lambda}}G_\lambda$
\begin{equation}\label{secondorder}
\partial^2_sQ(s) = -\Omega^2 Q(s) 
-\frac{1}{M} \sum_\lambda g_\lambda e^{-i\omega_\lambda s}
\langle z| U_t b_\lambda(s)|0\rangle
-\frac{1}{M} \sum_\lambda g_\lambda e^{i\omega_\lambda s}
\langle z| U_t b_\lambda^\dagger(s)|0\rangle. 
\end{equation}

As before in Section (\ref{heisenberg}), 
integrating the Heisenberg equation of motion for 
the environmental annihilation operator $b_\lambda(s)$, we see that
\begin{equation}\label{ingred1}
\langle z| U_t a_\lambda(s)|0\rangle =
-\frac{i}{\hbar}g_\lambda\int_0^s ds'\;e^{i\omega_\lambda s'}
\langle z| U_t q(s')|0\rangle,
\end{equation}
where use is made
of the fact that the initial 
$b_\lambda(0)|0\rangle = b_\lambda|0\rangle = 0$.

The adjoint $b_\lambda^\dagger(s)$ in (\ref{secondorder})
is dealt with in similar fashion.
Here, however, we make use of the fact that at the final time
$t$ the overall expression becomes simple: 
$\langle z| U_t b_\lambda^\dagger(t)|0\rangle
= z^*_\lambda \langle z| U_t|0\rangle$. Thus, we
integrate the Heisenberg equation of motion for $b_\lambda^\dagger(s)$
given the {\it final} value at $s=t$ to get
$b_\lambda^\dagger(s) = 
b_\lambda^\dagger(t) -\frac{i}{\hbar}
g_\lambda\int_s^t ds'\;e^{-i\omega_\lambda s'}
q(s')$, and find
\begin{equation}\label{ingred2}
\langle z| U_t b_\lambda^\dagger(s)|0\rangle =
z_\lambda^*\langle z| U_t|0\rangle
-\frac{i}{\hbar}g_\lambda\int_s^t ds'\;e^{-i\omega_\lambda s'}
\langle z| U_t q(s')|0\rangle.
\end{equation}

All that remains to be done is combining the results
(\ref{secondorder}), (\ref{ingred1}), and (\ref{ingred2}) 
to obtain a
linear second order differential equation for $Q(s)$:
\begin{eqnarray}\label{firstQBMo}
& & \partial^2_s Q(s) + \Omega^2 Q(s) \\ \nonumber
& & -\frac{i}{M\hbar}\int_0^s ds' \; \alpha(s-s') Q(s')
-\frac{i}{M\hbar}\int_s^t ds' \; \alpha(s'-s) Q(s') 
= -\frac{i}{M} z_s^* \langle z| U_t|0\rangle,
\end{eqnarray}
for $s\in [0,t]$ with a fixed final time $t$, and
final values (\ref{finalAs}).

With expression (\ref{Aop2}),
the derived second order differential equation for $Q(s)$
translates into an equation for the desired
operator $O(t,s,z^*)$.
Considered as a function of the time $s$,
at fixed $t$, we find
\begin{eqnarray}\label{QBMos}
& & \partial^2_s O(t,s,z^*) + \Omega^2 O(t,s,z^*)  \\ \nonumber
& & -\frac{i}{M\hbar}\int_0^s ds' \; \alpha(s-s') O(t,s',z^*)
-\frac{i}{M\hbar}\int_s^t ds' \; \alpha(s'-s) O(t,s',z^*) 
= -\frac{i}{M\hbar} z_s^* 
\end{eqnarray}
with final values from (\ref{finalAs})
\begin{equation}\label{ofinals}
O(t,s=t,z^*) = \frac{1}{\hbar}\,q,\;\;\mbox{and}\;\;
\left.\partial_sO(t,s,z^*)\right|_{s=t} = \frac{1}{M\hbar}\, p.
\end{equation}

As before in (\ref{Oharmonic}), 
we write the operator $O(t,s,z^*)$ in the form
\begin{equation}\label{QBMo}
O(t,s,z^*) =\frac{1}{\hbar}\left( 
\phi_t(s) q +\frac{1}{M\Omega} \psi_t(s) p 
- \frac{i}{M\Omega}\int_0^t ds' \chi_t(s,s') z_{s'}^*\right),
\end{equation}
where we introduced three $s$-dependent complex functions 
$\phi_t(s), \psi_t(s)$
and $\chi_t(s,s')$ at a fixed time $t$.
Obviously, from (\ref{QBMos}), the first two of them
satisfy the homogeneous equation
\begin{equation}\label{phipsi}
\partial_s^2 \phi_t(s) + \Omega^2 \phi_t(s) 
-\frac{i}{M\hbar}\int_0^s ds'' \; \alpha(s-s'') \phi_t(s'')
-\frac{i}{M\hbar}\int_s^t ds'' \; \alpha(s''-s) \phi_t(s'') = 0.
\end{equation}
While for $\phi_t$ we demand the final values 
\begin{equation}\label{condphi}
\phi_t(s=t) = 1\;\;\;\mbox{and}\;\;\;
\left.\partial_s\phi_t(s)\right|_{s=t} = 0
\end{equation}
to satisfy (\ref{ofinals}),
$\psi_t(s)$ has
final values 
\begin{equation}\label{condpsi}
\psi_t(s=t) = 0\;\;\;\mbox{and}\;\;\;
\left.\partial_s\psi_t(s)\right|_{s=t} = \Omega.
\end{equation}

The function $\chi_t(s,s,')$ in (\ref{QBMo}) is nothing but
Green's function that satisfies
\begin{eqnarray}\label{chi}
& & \partial_s^2 \chi_t(s,s') + \Omega^2 \chi_t(s,s') \\ \nonumber
& & -\frac{i}{M\hbar}\int_0^s ds'' \; \alpha(s-s'') \chi_t(s'',s')
-\frac{i}{M\hbar}\int_s^t ds'' \; \alpha(s''-s) \chi_t(s'',s') 
= \Omega\, \delta(s-s')
\end{eqnarray}
with final values 
\begin{equation}\label{condchi}
\chi_t(s=t,s') = 0\;\;\;\mbox{and}\;\;\;
\left.\partial_s\chi_t(s,s')\right|_{s=t} = 0
\end{equation}
for all $s' \in [0,t]$.

The result (\ref{QBMo}) for the operator $O(t,s,z^*)$ reflects
our ansatz (\ref{Oharmonic}) that solved
evolution equation (\ref{Oopevo}) with respect to $t$.
Clearly, we identify the relations
$f(t,s) = \phi_t(s)$, $g(t,s) = \psi_t(s)$, and
$j(t,s,s') = \chi_t(s,s')$.
All conditions (\ref{boundaries})
on the functions $f,g,j$ are met by 
(\ref{condphi}), (\ref{condpsi}), and (\ref{condchi}). One may
even derive their evolution equations (\ref{fgjs}) regarded as
a function of $t$ from the corresponding evolution
equations (\ref{phipsi}) and (\ref{chi}), when
considered as functions of $s$, see appendix \ref{appendixA}
for an example.
For numerical applications as in \cite{Str01},
the evolution equations with respect to $t$ are the
ones of interest. In order to derive the convolutionless
master equation for this quantum Brownian motion
model, however, it turns out that
the evolution equation of $O(t,s,z^*)$ with respect to
$s$ from (\ref{QBMos}) is the right starting point, as 
will be explained in the next section.

\section{QBM master equation}\label{sqbm2}

In the last Section we have established the evolution equation
with respect to $s$ for the operator $O(t,s,z^*)$ 
entering the exact convolutionless stochastic Schr\"odinger 
equation (\ref{qbmosci}).
We are now in the
position to derive the 
corresponding convolutionless master equation
from the general expression (\ref{rhoevo2}),
\begin{equation}\label{evorho2q}
\partial_t\rho_t = -\frac{i}{\hbar}[H,\rho_t]
+\frac{1}{\hbar}
[q,{\cal M}\left\{P_t {\bar O}^\dagger(t,z^*)\right\}] + 
\frac{1}{\hbar}[{\cal M}\left\{{\bar
O}(t,z^*)P_t\right\},q]
\end{equation}
which for the Quantum Brownian motion case
we display here with coupling operator $L=q/\hbar$.

We saw previously how knowledge
about the operator $O(t,s,z^*)$ replacing the
functional derivative in our stochastic Schr\"odinger
equation, puts us in a position
to derive a closed convolutionless
master equation. In particular,
as soon as the operator $O(t,s,z^*)$
is independent of the noise, we get
${\cal M}\{{\bar O}(t)P_t\}= {\bar O}(t)\rho_t$
and (\ref{evorho2q}) provides the convolutionless
master equation (\ref{evorho3}) for $\rho_t$.
In our current example, however, ${\bar O}(t,z^*)$ {\it does}
depend on the noise and we show next how to
proceed in this case.

Starting point is the evolution equation (\ref{QBMos}) for the
operator $O(t,s,z^*)$ {\it as a function of} $s$.
For the master equation (\ref{evorho2q}), we need an expression
for the ensemble mean 
\begin{equation}\label{ensmean22}
{\cal M}\left\{{\bar O}(t,z^*)P_t\right\} =
\int_0^t\,ds\, \alpha(t-s) {\cal M}\{O(t,s,z^*)P_t\} =
\int_0^t\,ds\, \alpha(t-s) R(t,s)
\end{equation}
with 
\begin{equation}\label{ensmean2}
R(t,s)\equiv {\cal M}\{O(t,s,z^*)P_t\}.
\end{equation}
Upon taking the mean ${\cal M}\{\ldots P_t\}$ of equation (\ref{QBMos}),
we see from its definition (\ref{ensmean2})
that $R(t,s)$ satisfies the second order differential equation
\begin{eqnarray}\label{Req1}
& & \partial^2_sR(t,s) + \Omega^2 R(t,s)\\
\nonumber & &  -\frac{i}{M\hbar}\int_0^s ds'\;\alpha(s-s')R(t,s')
-\frac{i}{M\hbar}\int_s^t ds'\;\alpha(s'-s) R(t,s') = 
-\frac{i}{M\hbar}{\cal M}\{z_s^* P_t\}.
\end{eqnarray}

Next we have to find an expression for ${\cal M}\{z_s^* P_t\}$, which
again follows from Novikov's theorem \cite{novikov} 
as in (\ref{novi}). We may
express the resulting functional derivative
in terms of the operator $O(t,s,z^*)$ and get
\begin{equation}\label{meanzP}
{\cal M}\{z_s^* P_t\} = \int_0^t ds' \alpha^*(s-s') 
{\cal M}\{P_t O^\dagger(t,s',z^*)\}
 = \int_0^t ds' \alpha^*(s-s') R^\dagger(t,s').
\end{equation}

A closed equation for the operator $R(t,s)$
as a function of $s$ results:
\begin{eqnarray}\label{Req2}
\partial_s^2 R(t,s) +& &  \Omega^2 R(t,s)
-\frac{i}{M\hbar}\int_0^s ds'\;\alpha(s-s')R(t,s') \\ \nonumber
& & 
-\frac{i}{M\hbar}\int_s^t ds'\;\alpha(s'-s) R(t,s') =
 -\frac{i}{M\hbar} \int_0^t ds' \alpha^*(s-s') R^\dagger(t,s').
\end{eqnarray}
Required final values are
\begin{equation}\label{Rfinals}
R(t,s=t) =\frac{1}{\hbar} q\rho_t,\;\;\mbox{and}
\;\; \left.\partial_s R(t,s)\right|_{s=t} =\frac{1}{M\hbar} p\rho_t,
\end{equation}
as one can easily see from (\ref{ofinals}).

In fact, it is clear that equation (\ref{Req2}) with
final values (\ref{Rfinals}) may be satisfied by an
expression of the form
\begin{equation}\label{Rexp}
R(t,s) =\frac{1}{\hbar}\left\{ 
k(t,s)q\rho_t + 
\frac{1}{M\Omega} \ell(t,s) p\rho_t + m(t,s) \rho_t q +
\frac{1}{M\Omega}  n(t,s) \rho_t p
\right\},
\end{equation}
where $k(t,s), \ell(t,s), m(t,s),$ and $n(t,s)$ are complex functions
whose $s$-dependence follows from equation (\ref{Req2})
with appropriate final values at $s=t$ from (\ref{Rfinals}).
As an example, we get $k(t,s=t)=1$ and $\partial_s k(t,s)|_{s=t} = 0$.
In Appendix \ref{appendixB} we show
how to determine all four functions.

Now that we know $R(t,s)=  {\cal M}\{O(t,s,z^*)P_t\}$ from
(\ref{Rexp}),
the closed master equation for the reduced density 
operator follows immediately from the general
expression (\ref{evorho2q}) with (\ref{ensmean22}).
After some rearrangements, we indeed arrive at the more familiar form
(\ref{masterosci}), with
time dependent coefficients
\begin{eqnarray}\label{qbmfunctions}
a(t) & = & \frac{2}{\hbar}
\mbox{Im}\left(\int_0^t ds\, \alpha(t-s) (k(t,s)+m(t,s))\right)
\\ \nonumber
b(t) & = & \frac{2}{M\Omega\hbar}
\mbox{Im}\left(\int_0^t ds\, \alpha(t-s) (\ell(t,s)+n(t,s))\right)
\\ \nonumber
c(t) & = & \frac{1}{M\Omega} 
\mbox{Re}\left(\int_0^t ds\, \alpha(t-s) (n(t,s)-\ell(t,s))\right)
\\ \nonumber
d(t) & = & \mbox{Re}\left(\int_0^t ds\, \alpha(t-s) (k(t,s)-m(t,s))\right).
\end{eqnarray}
In appendix \ref{appendixB} we discuss these expressions in more
detail, establishing the connection to earlier derivations of the
master equation (\ref{masterosci}).  Crucially, we here show
explicitely that it follows from the corresponding 
convolutionless stochastic Schr\"odinger equation (\ref{qbmosci}).

\section{Conclusion}\label{conclusion}

Exactly soluble models are valuable tools. They allow us to discuss
approximation schemes, show the transitions in the qualitative
behaviour when parameters are changed over a wide range, and last 
not least provide insight into different theoretical approaches.

Our non-Markovian stochastic Schr\"odinger equation approach offers 
a new method for handling open quantum systems.
In this paper we use this framework as a theoretical tool
to derive convolutionless non-Markovian master equations.
Most interestingly, the
exact master equation for quantum Brownian motion of a harmonic
oscillator, coupled to a bath of oscillators is derived from
the corresponding convolutionless 
non-Markovian stochastic Schr\"odinger equation.
In both approaches, non-Markovian
properties are encoded in time-dependent coefficients.

The problem how to describe non-Markovian open quantum system dynamics
efficiently is a difficult one. As further underlined in this papers,
we believe that our non-Markovian stochastic Schr\"odinger equation 
approach offers a useful alternative framework to the existing ones.

\section*{Acknowledgments}
We wish to thank Joe  Eberly, Nicolas Gisin, Fritz Haake and Ian Percival for valuable 
discussions.
Part of this work was completed during a visit of WTS at Queen Mary College,
supported by the ``Nachkontaktprogramm'' of the Alexander von
Humboldt-Foundation. WTS also thanks
the Deutsche Forschungsgemeinschaft for support through the SFB 276
``Korrelierte Dynamik hochangeregter atomarer und molekularer
Systeme''. TY thanks the Leverhulme
Foundation and a grant from the NEC Research Institute for support in London
and Rochester respectively.

\begin{appendix}

\section{Evolution equation with respect to time $t$}\label{appendixA}

It is interesting to see the connection between
the evolution equations
(\ref{fgjs}) with respect to $t$
for $f(t,s)=\phi_t(s), g(t,s)=\psi_t(s)$
and $j(t,s,s')=\chi_t(s,s')$ and the
differential equations (\ref{phipsi}) and (\ref{chi}) 
for the same functions with respect to $s$.
In fact, here we show how to derive the former from the latter.

Consider $g(t,s)=\psi_t(s)$ as an example. We showed in Section 
\ref{heisenbergQBM} that this function solves (\ref{phipsi}) which we here
display again for convenience:
\begin{equation}\label{phipsi2}
\partial_s^2 \psi_t(s) + \Omega^2 \psi_t(s) 
-\frac{i}{M\hbar}\int_0^s ds'' \; \alpha(s-s'') \psi_t(s'')
-\frac{i}{M\hbar}\int_s^t ds'' \; \alpha(s''-s) \psi_t(s'') = 0.
\end{equation}

Taking the time derivative of this equation with respect to $t$,
and using the final condition $\psi_t(s=t)=0$ from (\ref{condpsi}), 
we find that $\partial_t\psi_t(s)$ satisfies the very same
equation (\ref{phipsi2}). Therefore,
as $\{\phi_t(s),\psi_t(s)\}$ forms a basis of solutions, there exists a
relation of the form
\begin{equation}\label{phievo}
\partial_t\psi_t(s) = c_1(t)\phi_t(s) + c_2(t)\psi_t(s)
\end{equation}
with suitable coefficients $c_1(t)$ and $c_2(t)$. These have to be 
determined from
the final values at $s=t$. From (\ref{condphi}) and
(\ref{condpsi}) we get
\begin{eqnarray}\label{coefs}
c_1(t) & = & \partial_t \psi_t(s=t) \\ \nonumber
c_2(t) & = &\left.\frac{1}{\Omega}\partial_t \partial_s\psi_t(s)\right|_{s=t}.
\end{eqnarray}

These expressions are easily evaluated: first, from $\psi_t(s=t)=0$,
we get $\partial_t\psi_t(s=t) + \left.\partial_s\psi_t(s)\right|_{s=t} = 0$
and thus $c_1(t)= \left.-\partial_s\psi_t(s)\right|_{s=t} = -\Omega$
according to (\ref{condpsi}). Secondly, we use the trivial
identity 
\begin{eqnarray}
\partial_s\psi_t(s) & = &
\left.\partial_s\psi_t(s)\right|_{s=t} -\int_s^t\,ds''\,
\left(\left.\partial^2_s\psi_t(s)\right|_{s=s''}\right) \\ \nonumber
& = & \Omega -\int_s^t\,ds''\,\left(
\left.\partial^2_s\psi_t(s)\right|_{s=s''}\right).
\end{eqnarray}
Take the derivative with respect to $t$ and then set $s=t$ to find
\begin{eqnarray}
\left.\partial_t\partial_s\psi_t(s)\right|_{s=t} & = &
-\left.\partial^2_s\psi_t(s)\right|_{s=t} \\ \nonumber
& = & \Omega^2\psi_t(s=t)
-\frac{i}{M\hbar}\int_0^t\, ds'' \; \alpha(t-s'') \psi_t(s'')\\ \nonumber
& = & -\frac{i}{M\hbar}\int_0^t\, ds'' \; \alpha(t-s'') \psi_t(s''),
\end{eqnarray}
where use is made of the evolution equation (\ref{phipsi2}) and
the condition (\ref{condpsi}). Thus, with (\ref{coefs}),
$c_2(t) =  -\frac{i}{M\Omega\hbar}\int_0^t\, ds'' \; \alpha(t-s'') \psi_t(s'')$
which may be written as 
$c_2(t) =  -i G(t)$ using definition (\ref{FGJs}).
We see that with $c_1(t)=-\Omega$ and
$c_2(t) =  -i G(t)$, relation (\ref{phievo}) is nothing but
the evolution equation for $\psi_t(s)=g(t,s)$ in (\ref{fgjs}) as derived 
from the ``consistency condition'' (\ref{consistency}).

\section{Time dependent coefficients}\label{appendixB}

From the second order differential equation (\ref{Req2})
for the operator $R(t,s)$ 
which we here display again for convenience,
\begin{equation}\label{Rexpapp}
R(t,s) =\frac{1}{\hbar}\left\{ 
k(t,s)q\rho_t + 
\frac{1}{M\Omega} \ell(t,s) p\rho_t + m(t,s) \rho_t q +
\frac{1}{M\Omega}  n(t,s) \rho_t p
\right\},
\end{equation}
we get (coupled) differential equations for the coefficients 
$k(t,s), \ell(t,s), m(t,s),$ and $n(t,s)$ as functions of $s$.
The final values at $s=t$ for all four functions
are determined from the
conditions (\ref{Rfinals}) on $R(t,s)$.

It turns out useful to replace the four unknown functions
by the linear combinations
appearing in the final result (\ref{qbmfunctions}), which we define as
\begin{eqnarray}\label{XYZT}
w(t,s) & \equiv &  k(t,s) + m(t,s) \\ \nonumber
x(t,s) & \equiv &  \ell(t,s) + n(t,s) \\ \nonumber
y(t,s) & \equiv &  k(t,s) - m(t,s) \\ \nonumber
z(t,s) & \equiv &  n(t,s) - \ell(t,s)
\end{eqnarray}
They all satisfy {\it uncoupled} second order differential equations.
For $w(t,s)$ we find from (\ref{Req2})
\begin{eqnarray}\label{WXeqs}
&&\partial^2_s w(t,s) + \Omega^2 w(t,s) -\frac{i}{M\hbar}
\int_0^s ds' \alpha(s-s')
w(t,s')\nonumber\\
&&-\frac{i}{M\hbar}\int_s^t\alpha(s'-s) w(t,s') 
=-\frac{i}{M\hbar} \int_0^t ds' \alpha(s'-s) w^*(s'),
\end{eqnarray}
and the identical equation for $x(t,s)$. Required final values are
\begin{eqnarray}\label{wxfinals}
w(t,s=t) = 1,\;\; & & \partial_s w(t,s)|_{s=t} = 0 \\ \nonumber
x(t,s=t) = 0,\;\; & & \partial_s x(t,s)|_{s=t} = \Omega.
\end{eqnarray}
For $y(t,s)$ (and also $z(t,s)$)
we get a similar equation with a different sign on the right hand side:
\begin{eqnarray}\label{YZeqs}
&&\partial^2_s y(t,s) + \Omega^2 y(t,s) -\frac{i}{M\hbar}
\int_0^s ds' \alpha(s-s')
y(t,s')\nonumber\\
&&-\frac{i}{M\hbar}\int_s^t\alpha(s'-s) y(t,s') 
=\frac{i}{M\hbar} \int_0^t ds' \alpha(s'-s) y^*(s'),
\end{eqnarray}
here with final values
\begin{eqnarray}\label{yzfinals}
y(t,s=t) = 1,\;\; & & \partial_s y(t,s)|_{s=t} = 0 \\ \nonumber
z(t,s=t) = 0,\;\; & & \partial_s z(t,s)|_{s=t} = -\Omega.
\end{eqnarray}

It turns out that matters simplify once we separate
real- and imaginary parts of the four functions $w,x,y,z$.
For the bath correlation function we write
\begin{equation}\label{alphareim}
\alpha(t-s) = \nu(t-s) + i \hbar\eta(t-s).
\end{equation}
Thus, microscopically, from its definition (\ref{alpha})
we have
\begin{equation}
\eta(t-s) = -\sum_\lambda\frac{G_\lambda^2}{2m_\lambda\omega_\lambda}
\sin\omega_\lambda(t-s) = -\int_0^\infty d\omega J(\omega)\sin\omega (t-s).
\end{equation}
This kernel turns out to be crucial for the {\it classical} equation of
motion of the underlying model (\ref{qbmH}), as will be clarified
shortly.

For the imaginary part of $w(t,s)$ we get from (\ref{WXeqs})
\begin{equation}\label{WI}
\partial_s^2 w_I(t,s) + \Omega^2 w_I(t,s) +\frac{2}{M}\int_s^t ds'
\eta(s-s')w_I(t,s') = 0
\end{equation}
and the very same equation for the imaginary part of $x(t,s)$.
With final values $w_I(t,t) = 0, \partial_s w_I(t,s)|_{s=t} = 0$ and
$x_I(t,t) = 0, \partial_s x_I(t,s)|_{s=t} = 0$
the solutions are trivial:
\begin{eqnarray}\label{WXIfinal}
w_I(t,s) & \equiv & 0 \;\;\mbox{for all}\;\; s\\ \nonumber
x_I(t,s) & \equiv & 0 \;\;\mbox{for all}\;\; s.
\end{eqnarray}
Next we consider the real parts $w_R(t,s), x_R(t,s)$. With
(\ref{WXIfinal}) and (\ref{WXeqs}) we arrive at
\begin{equation}\label{WXR}
\partial_s^2 w_R(t,s) + \Omega^2 w_R(t,s) + \frac{2}{M}\int_0^s ds'
\eta(s-s')w_R(t,s') = 0
\end{equation}
and the same equation for $x_R(t,s)$.
Required final values are
\begin{eqnarray}\label{wxRfinals}
w_R(t,s=t) = 1,\;\; & & \partial_s w_R(t,s)|_{s=t} = 0 \\ \nonumber
x_R(t,s=t) = 0,\;\; & & \partial_s x_R(t,s)|_{s=t} = \Omega.
\end{eqnarray}

Similarly, we find for the real and imaginary
part of the function $y(t,s)$ 
the following equations:
\begin{eqnarray}\label{YZRIeqs}
\partial_s^2 y_R(t,s) + \Omega^2 y_R(t,s) + \frac{2}{M}
\int_s^t ds'\eta(s'-s) y_R(t,s') 
& = & 0\\ \nonumber
\partial_s^2 y_I(t,s) + \Omega^2 y_I(t,s) + \frac{2}{M}
\int_0^s ds'\eta(s-s') y_I(t,s') 
& = & \frac{2}{M\hbar}\int_0^tds'\nu(s-s')y_R(t,s').
\end{eqnarray}
As before, real and imaginary part of $z(t,s)$ satisfy the very same equations.
Required ``final values'' are
\begin{eqnarray}\label{YZfinals}
y_R(t,t) = 1,\;\; & & \partial_s y_R(t,s)|_{s=t} = 0 \\ \nonumber
y_I(t,t) = 0,\;\; & & \partial_s y_I(t,s)|_{s=t} = 0 \\ \nonumber
z_R(t,t) = 0,\;\; & & \partial_s z_R(t,s)|_{s=t} = -\Omega \\ \nonumber
z_I(t,t) = 0,\;\; & & \partial_s z_I(t,s)|_{s=t} = 0.
\end{eqnarray}

The real parts of all four functions may nicely be expressed
using the special solution $q(s)$ of equation (\ref{WXR})
\begin{equation}\label{qs}
\partial_s^2 q(s) + \Omega^2 q(s) + \frac{2}{M}\int_0^s ds'
\eta(s-s')q(s') = 0
\end{equation}
with {\it initial} values 
\begin{equation}\label{qinitial}
q(0) = 0,\;\;\mbox{and}\;\;  \partial_s q(s)|_{s=0} = \Omega.
\end{equation}
Equation (\ref{qs}) is nothing but the {\it classical} equation
of motion for the position $q(s)$ of the underlying model (\ref{qbmH}),
provided the environmental oscillators all start with initial
values $q_\lambda(0) = 0,\; {\dot q}_\lambda(0) = 0$.
Note that ${\dot q}(s)$ satisfies the very same
equation (\ref{qs}) with ${\dot q}(0)=\Omega$ and ${\ddot q}(0)=0$.

It is customary to write equation (\ref{qs}) in
the more familiar form
\begin{equation}\label{qs2}
\partial_s^2 q(s) + \Omega^2 q(s) + \int_0^s ds'
\gamma_{\rm cl}(s-s'){\dot q}(s') = 0
\end{equation}
with the classical damping kernel $\gamma_{\rm cl}(s-s')$
defined through
$\frac{2}{M}\eta(t-s) \equiv \partial_t \gamma_{\rm cl}(t-s)$.

In terms of this classical solution $q(s)$ simple
inspection shows that
\begin{eqnarray}\label{finalexpR}
w(t,s) = w_R(t,s) & = & \frac{{\dot q}(s){\dot q}(t) - q(s){\ddot q}(t)}
                             {({\dot q}(t))^2 - q(t){\ddot q}(t)},\\ \nonumber
x(t,s) = x_R(t,s) & = & \Omega
                           \frac{q(s){\dot q}(t) - {\dot q}(s)q(t)}
                             {({\dot q}(t))^2 - q(t){\ddot q}(t)},\\ \nonumber
y_R(t,s) & = & \frac{1}{\Omega} {\dot q}(t-s) ,\\ \nonumber
z_R(t,s) & = & q(t-s).
\end{eqnarray}

While the imaginary parts of $w(t,s)$ and $x(t,s)$ are zero, the
expressions for $y_I(t,s)$ and $z_I(t,s)$ are more involved.
We find
\begin{eqnarray}\label{finalexpI}
y_I(t,s) & = & 
   \frac{2}{M\Omega^2\hbar}
\int_0^sds'\int_0^tds'' \nu(s'-s''){\dot q}(t-s'')q(s-s')\\ \nonumber
 & & -\frac{2}{M\Omega^2\hbar}
\int_0^tds'\int_0^tds'' \nu(s'-s''){\dot q}(s'')
    [w(t,s)q(s')+x(t,s){\dot q}(s')/\Omega],
\\ \nonumber
z_I(t,s) & = & 
   \frac{2}{M\Omega\hbar}
\int_0^sds'\int_0^tds'' \nu(s'-s'')q(t-s'')q(s-s')\\ \nonumber
 & & -\frac{2}{M\Omega\hbar}
\int_0^tds'\int_0^tds'' \nu(s'-s'')q(s'')
    [w(t,s)q(s')+x(t,s){\dot q}(s')/\Omega].
\end{eqnarray}
The first line in each expression is simply the inhomogeneity
on the right hand side of equations (\ref{YZRIeqs}), in convolution with
Green's function $q(s-s')/\Omega$. The second lines, respectively,
are solutions of the homogeneous equations and
serve to satisfy the final conditions (\ref{YZfinals}).

Finally, let us make contact to earlier derivations of the time
dependent coefficients. For brevity we concentrate on the drift
coefficients $a(t)$ and $b(t)$ in the master equation (\ref{masterosci}).
As $w(t,s)$ and $x(t,s)$ are real, using (\ref{qbmfunctions})
and (\ref{alphareim}), 
$a(t)$ and $b(t)$
may be written in the form
\begin{eqnarray}\label{qbmfunctions2}
a(t) & = & 2 \int_0^t ds\, \eta(t-s) w_R(t,s)
\\ \nonumber
b(t) & = & \frac{2}{M\Omega} \int_0^t ds\, \eta(t-s) x_R(t,s).
\end{eqnarray}

With expressions (\ref{finalexpR}), and using the evolution equation
(\ref{qs}) for both, $q(s)$ and ${\dot q}(s)$, we may write the drift
coefficients in the form
\begin{eqnarray}\label{qbmfunctions3}
a(t) & = & -M\Omega^2 + M\frac{({\ddot q}(t))^2 - {\dot q}(t){\dddot q}(t)}
                             {({\dot q}(t))^2 - q(t){\ddot q}(t)},
\\ \nonumber
b(t) & = & \frac{ q(t){\dddot q}(t)- {\dot q}(t){\ddot q}(t)}
                             {({\dot q}(t))^2 - q(t){\ddot q}(t)}.
\end{eqnarray}
Expressions similar to these may also be found in the 
literature \cite{HR,HPZ,HY96,CF}.

\end{appendix}

\end{document}